\documentclass[prb,twocolumn,showpacs,amsmath,amssymb]{revtex4}
\usepackage{graphicx}
\begin{document}
\newcommand{\bn}{{\bf n}}
\newcommand{\bp}{{\bf p}}
\newcommand{\br}{{\bf r}}
\newcommand{\bq}{{\bf q}}
\newcommand{\bj}{{\bf j}}
\newcommand{\eps}{\varepsilon}
\newcommand{\la}{\langle}
\newcommand{\ra}{\rangle}
\newcommand{\hx}{\hat x}
\newcommand{\hq}{\hat q}
\newcommand{\hp}{\hat p}
\newcommand{\hH}{{\hat H}_0}
\newcommand{\kw}{k_{\omega}}
\newcommand{\qw}{q_{\omega}}
\input epsf

\title{ Frequency dependent third cumulant of current in diffusive conductors }

\author{ S. Pilgram$^1$, K. E. Nagaev$^2$, and  M. B\"uttiker$^1$}
\affiliation{ $^1$D\'epartement de Physique Th\'eorique,
  Universit\'e de
  Gen\'eve, CH-1211, Gen\`eve 4, Switzerland\\
  $^2$Institute of Radioengineering and Electronics,
  Russian Academy of Sciences, Mokhovaya ulica 11, 125009 Moscow,
  Russia }

\date{\today}

\begin{abstract}
We calculate the frequency dispersion of the third cumulant of
current in diffusive-metal contacts. The cumulant exhibits a
dispersion at the inverse time of diffusion across the contact,
which is typically much smaller than the inverse $RC$ time. This
dispersion is much more pronounced in the case of  strong
electron-electron scattering than in the case of purely elastic
scattering because of a different symmetry of the relevant
second-order correlation functions.
\end{abstract}

\maketitle

\section{ Introduction }

Measurements of nonequilibrium noise provide valuable information
about the properties of a system, which cannot be extracted from
measurements of average quantities. For example, measurements of
shot noise give the magnitude of the quasiparticle charge in the
case of a tunnel contact and the effective temperature of
electrons in the case of a diffusive contact.\cite{Blanter-00}
Recently, Reulet et al.\cite{Reulet-03} performed first
measurements of the third cumulant of current, which may give even
more interesting information. For example, this cumulant is very
sensitive to the presence of electron-electron scattering in a
diffusive contact. Electron-electron scattering changes the shot
noise in a diffusive contact only by several
percent,\cite{Nagaev-95,Kozub-95} but it changes the third
cumulant of current almost by an order of
magnitude.\cite{Gutman-03}

Of special interest is the frequency dependence of the third
cumulant. Very recently, it was shown that third cumulants of
current in a chaotic cavity whose contacts have different
transparencies may exhibit a frequency dispersion much more
complicated than that of the shot noise. Unlike the conventional
shot noise that has a dispersion only at the inverse {\it $RC$ time} of
the cavity,\cite{Pedersen-98} the third cumulant of noise may also
exhibit a dispersion at the inverse {\it dwell time} of an electron on
the cavity.\cite{Nagaev-03} In most cases, this time is much
longer than the $RC$ time that describes charge relaxation in the
cavity, and therefore the corresponding dispersion takes place at
experimentally accessible frequencies. This dispersion is due to
slow fluctuations of the distribution function that do not violate
electroneutrality and are akin to fluctuations of local
temperature. These fluctuations do not directly contribute to the
current and therefore are not seen in conventional noise, but they
modulate the intensity of noise sources and therefore manifest
themselves in higher correlations of current.
\begin{figure}[t]
 \epsfxsize=8.5cm
 \epsfbox{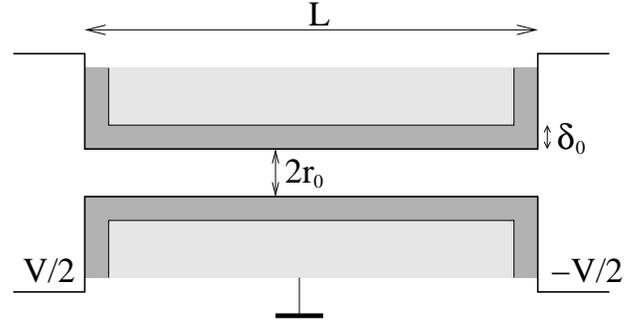}
 \caption{\label{fig:wire} The geometry of the system considered. White areas show the diffusive metal and dark areas,
 the insulator.}
\end{figure}

Another important example of a system with a long dwell time is a
diffusive contact. In this work we investigate the frequency
dependence of the third cumulant of a metallic diffusive wire.
Like a chaotic cavity, it also has a long dwell time. In addition,
the metallic diffusive wire is of interest because its impedance
can be easily matched to that required by current experimental
detection schemes.\cite{Reulet-03} Furthermore, the measuring
frequencies in this case are in the range where the frequency
dispersion takes place for the system at hand.

The zero-frequency third cumulant of current for a diffusive wire
was first calculated by Lee, Levitov, and Yakovets\cite{Lee-95}
for non-interacting electrons in the zero-temperature limit.
Recently this calculation has been extended to finite temperatures
and to the case of strong electron-electron scattering by Gutman
and Gefen.\cite{Gutman-03} In this paper we calculate the
frequency dependence of this quantity both for the case of
non-interacting electrons and for the hot-electron regime and show
that the latter case is more convenient for the experimental observation of
this effect, since the dispersion of the third cumulant is much stronger.

We present  a calculation based on the cascaded
Boltzmann--Langevin approach.\cite{Nagaev-02} In the Appendix, we
also derive the full generating functional for the frequency
dependent current fluctuations of a metallic wire both for the
elastic and hot-electron regime based on the stochastic
path-integral approach to full counting
statistics.\cite{Pilgram-03} The third cumulant of current may be
expressed in terms of functional derivatives of this functional.

\section{ Model and basic equations }

Consider a quasi-one-dimensional diffusive wire of length $L$ and
conductivity $\sigma$. To explicitly describe its electric
environment, the wire is chosen in the shape of a cylinder with a
diameter $2r_0$ and is embedded in a perfectly grounded medium,
which is separated from the wire by a thin insulating film of
thickness $\delta_0$ and with a dielectric constant $\eps_d$ (see
Fig.~1). All dimensions are assumed to be much larger than the
elastic mean free path and the screening length in the metal. The
electrodes are assumed to be perfect conductors, so the third
cumulant of current is not affected by the external
circuit.\cite{Beenakker-03,Kindermann-03,footnote} We also
restrict ourselves to sufficiently high voltages or temperatures,
hence the quantum dispersion of this quantity does not show up in
the frequency range of interest.\cite{Galaktionov-03} We emphasize
that despite the particular choice of geometry, our results are
valid for any quasi-one-dimensional diffusive contact.

With the above assumptions, the noise of current may be described
using the semiclassical Boltzmann--Langevin
approach.\cite{Kogan-69} The frequency dependence of shot noise in
diffusive contacts with account taken of electrical screening was
calculated in Refs. \onlinecite{Naveh-97, Nagaev-98}. To calculate
the frequency dependence of the third cumulant of current, we use
the cascade extension of this approach.\cite{Nagaev-02} The key
idea of this extension is a large separation between the time
scales describing the individual scattering events and the
evolution of the distribution function of electrons in the
contact. The resulting expressions may be also obtained by
considering the corresponding stochastic path
integral\cite{Pilgram-03,broader} for the diffusive
Boltzmann-Langevin equation (see Appendix). The cascade expansion
corresponds to a systematic expansion of the saddle-point
equations of this path integral in powers of the counting field.

The quantity we are going to calculate is the Fourier transform of the third
order current correlation function defined as
\begin{eqnarray}
 P_3(\omega_1,\omega_2)
 =
 \int d(t_1 - t_2) \int d(t_2 - t_3)
\nonumber\\
 \times
 \exp
 [
  i\omega_1(t_1 - t_3)
  +
  i\omega_2(t_2 - t_3)
 ]
 \la
  \delta I(t_1) \delta I(t_2) \delta I(t_3)
 \ra.
\label{def}
\end{eqnarray}
The starting point for our calculations is the stochastic
 diffusion equation
for the fluctuations $\delta f(\eps,\br)$ of the distribution
 function
$f(\eps,\br)$
\begin{equation}
 \left(
  \frac{\partial}{\partial t}
  -
  D\nabla^2
 \right)
 \delta f
 -
 \delta I_{ee}
 =
 -e
 \delta\dot{\phi}
 \frac{\partial f}{\partial\eps}
 -
 \nabla\delta{\bf F}^{imp}
 -
 \delta F^{ee},
 \label{kinetic}
\end{equation}
where $D$ is the diffusion coefficient, $\delta I_{ee}$ is the
linearized electron-electron collision integral, and $\delta{\bf
F}^{imp}$ and $\delta F^{ee}$ are random extraneous sources
associated with electron-impurity and electron-electron
scattering. This equation is obtained from the standard
Boltzmann--Langevin equation by defining the electron energy  as
$\eps = p^2/2m + e\phi(\br, t)- \eps_F$ and isolating the
isotropic part of the distribution function in the momentum space.
The fluctuation of the electric potential $\delta\phi$ that
appears in this equation should be calculated self-consistently
from the Poisson equation
\begin{equation}
 \nabla^2\delta\phi
 =
 -4\pi\delta\rho,
\label{Poisson}
\end{equation}
where the fluctuation of charge density $\delta\rho$ is given by
\begin{equation}
 \delta\rho
 =
 eN_F
 \left(
  \int d\eps \delta f(\eps)
  +
  e\delta\phi
 \right)
\label{self-consist}
\end{equation}
and where $N_F$ is the Fermi density of states. In the case of a
quasi-one-dimensional contact, a solution of Eqs. (\ref{kinetic}) -
(\ref{self-consist}) is of the form\cite{Nagaev-98}
\begin{eqnarray}
 \delta\phi(x,\omega)
 =
 \frac{1}{S_0 \sigma}
 \left(
  \nabla^2
  +
  i\omega RC/L^2
 \right)^{-1}
\\ \nonumber
 \times
 \frac{\partial}{\partial x}
 \int d^2r_{\perp}
 \delta j_x^{ext}(\br),
 \label{dphi}
\end{eqnarray}
where $x$ is the coordinate along the contact, $\sigma=e^2N_FD$ is the
conductivity of the metal, $S_0 = \pi r_0^2$ is the cross section area of the
contact, $C = L\eps_d r_0/2\delta_0$ and $R = L/\pi r_0^2\sigma$ are the
capacitance and the resistance of the contact, and
\begin{equation}
  \delta\bj^{ext}
  =
  eN_F
  \int d\eps
  \delta{\bf F}^{imp}.
  \label{dj-vs-df}
\end{equation}
A fluctuation of the total current density is given by
\begin{eqnarray}
 \delta {\bf j}
 =
 \delta {\bf j}^{ext}
 -
 \sigma \nabla \delta\phi,
 \label{tot_curr}
\end{eqnarray}
and a fluctuation of the total current at the left end of the contact thus
equals
\begin{equation}
 \delta I
 =
 \sigma
 \int d^2 r_{\perp}
 \left.
  \frac
  { \partial\delta\phi(x, \omega) }
  { \partial x }
 \right|_{x=-L/2}.
 \label{dI_L}
\end{equation}
Making use of the correlation function of extraneous sources
\begin{eqnarray}
 \la
   \delta F_{\alpha}^{imp}(\eps, \br)
   \delta F_{\beta }^{imp}(\eps', \br')
 \ra_{\omega}
 &=&
 2\frac{D}{N_F}
 \delta(\br - \br')
 \delta(\eps - \eps')
\nonumber\\
 \times
 \delta_{\alpha\beta}
 f(\eps,\br)[1 &-& f(\eps,\br)],
 \label{<dF^2>}
\end{eqnarray}
one easily obtains the second-order correlation function for the fluctuations
of the current as a functional of the distribution function 
$f$.\cite{Nagaev-92}

Consider now the third cumulant of current. As the direct
contribution to this quantity from the third cumulant of
extraneous sources is negligibly small in a diffusive
metal,\cite{Nagaev-02} this quantity is dominated by an indirect
contribution of the second cumulant of these sources, which
results from the modulation of their intensity by fluctuations of the
distribution function. It may be written in the form
\begin{equation}
  \la
    \delta I(t_1)
    \delta I(t_2)
    \delta I(t_3)
  \ra
  =
  {\cal P}_{123}
  \{
   \Delta_{123}
  \},
\label{permut}
\end{equation}
where
\begin{eqnarray}
  \Delta_{123}
  =
  \int dt \int d\eps \int d^3r\,
  \frac{
    \delta
    \la
      \delta I(t_1)
      \delta I(t_2)
    \ra
  }{
    \delta f(\eps, \br, t)
  }
\nonumber\\
  \times
  \la
    \delta f(\eps, \br, t)
    \delta I(t_3)
  \ra
  \label{I^3-gen}
\end{eqnarray}
and ${\cal P}_{123}$ denotes a summation over all inequivalent permutations of
indices (123).

Equations (\ref{dphi}) and (\ref{dI_L}) suggest that the second cumulant of
current exhibits a dispersion at frequencies of the order of
$(RC)^{-1}$. Typically such high frequencies are beyond the experimentally
accessible range. Therefore in what follows we will assume that the
frequencies $\omega_1$, $\omega_2$, and $\omega_3$ are much smaller than
$(RC)^{-1}$. Hence the pile-up of charge in the contact may be neglected and
the fluctuation of current may be considered as coordinate independent. In
this case,
\begin{eqnarray}
 \la
  \delta I(\omega_1)
  \delta I(\omega_2)
 \ra
 =
 4\pi
 \delta( \omega_1 + \omega_2 )
 (RL)^{-1}
\\ \nonumber
 \times
 \int dx \int d\eps
 f(\eps, x)[1 - f(\eps,x)]
\label{<I^2>}
\end{eqnarray}
and the only possible dispersion in Eq. (\ref{I^3-gen}) is due to
the dynamics of a fluctuation $\delta f$, so that the expression
for the third cumulant assumes the form
\begin{eqnarray}
  P_3(\omega_1,\omega_2)
  =
  P(\omega_1) + P(\omega_2) + P(-\omega_1 - \omega_2),
  \qquad
\nonumber\\
  P(\omega)
  =
  \frac{2}{RL}
  \int\limits_{-L/2}^{L/2} dx
  \int d\eps\,
  [
   1 - 2f(\eps, x)
  ]
   \la
     \delta f(\eps,x)
     \delta I
   \ra_{\omega}.\quad
  \label{P}
\end{eqnarray}
The quantity $P(\omega)$ has to be calculated in different ways
for the case of purely elastic scattering and for the hot-electron
regime.

\section{ Purely elastic scattering }

For purely elastic scattering, $\delta I^{ee}$ and $\delta F^{ee}$
in Eq.  (\ref{kinetic}) vanish, and a fluctuation of the distribution function
$\delta f$ may be presented as a sum of a part induced directly by an
extraneous source
\begin {equation}
 \delta f_F(\eps,x,\omega)
 =
 \left(
  \nabla^2
  +
  i\omega/D
 \right)^{-1}
 \frac{
  \partial F_x^{ext}
 }{
  \partial x
 }
\end{equation}
and a part induced by fluctuations of the electric potential
\begin{equation}
 \delta f_\phi(\eps,x,\omega)
 =
 -i\omega
 \left(
  \nabla^2
  +
  i\omega/D
 \right)^{-1}
 \left[
  \frac{
   \partial f(\eps, x)
  }{
   \partial\eps
  }
  e
  \delta\phi(x,\omega)
 \right].
\label{df2}
\end{equation}
The existence of the term (\ref{df2}) indicates that the dynamics
of charged electrons differs from the dynamics of neutral
particles even at frequencies much smaller than $(RC)^{-1}$.

By multiplying these equations with the fluctuation of current
(\ref{dI_L}) and making use of the correlation function
(\ref{<dF^2>}), we obtain
\begin{eqnarray}
 \la
  \delta f_F(\eps,x)
  \delta I
 \ra_{\omega}
 =
 -2
 \frac{e}{L}
 \left(
  \nabla^2
  -
  i\omega/D
 \right)^{-1}
\nonumber\\
 \times
 \frac{
  \partial
 }{
  \partial x
 }
 \left\{
  f(\eps,x)[1 - f(\eps,x)]
 \right\}
\label{<df1dI>}
\end{eqnarray}
and
\begin{eqnarray}
 \la
  \delta f_{\phi}(\eps,x)
  \delta I
 \ra_{\omega}
 =
 i\omega
 \left(
  \nabla^2
  -
  i\omega/D
 \right)^{-1}
\nonumber\\
 \times
 \left[
  \frac{
   \partial f(\eps,x)
  }{
   \partial\eps
  }
  e
  \la
   \delta\phi(x)
   \delta I
  \ra_{\omega}
 \right].
\label{<df2dI>}
\end{eqnarray}
At low frequencies, one easily obtains from Eqs. (\ref{dphi}) and (\ref{dI_L})
that
\begin{equation}
 \la
  \delta\phi(x)
  \delta I
 \ra_{\omega}
 =
 \frac{2}{L}
 (
  \nabla^2
 )^{-1}
 \left[
  \frac{\partial}{\partial x}
  \int d\eps\,
  f(1 - f)
 \right].
 \label{<dphidI>-zf}
\end{equation}
Using the well-known expression for the average distribution function
\begin{equation}
 \bar{f}(\eps,x)
 =
 \left(
  \frac{1}{2}
  +
  \frac{x}{L}
 \right)
 f_0(\eps + eV/2)
 +
 \left(
  \frac{1}{2}
  -
  \frac{x}{L}
 \right)
 f_0(\eps - eV/2),
\label{f-elast}
\end{equation}
where $f_0$ is the equilibrium Fermi distribution and $V$ is the
voltage drop across the contact, we obtain
\begin{equation}
 \la
  \delta\phi(x)
  \delta I
 \ra_{\omega}
 =
 \frac{1}{6}
 \frac{x}{L}
 \left(
  1 - 4\frac{x^2}{L^2}
 \right)
 \left[
  eV
  \coth
  \left(
   \frac{eV}{2T}
  \right)
  -
  2T
 \right].
\label{<dphidI>-elast}
\end{equation}
\begin{figure}
 \epsfxsize=8.5cm
 \epsfbox{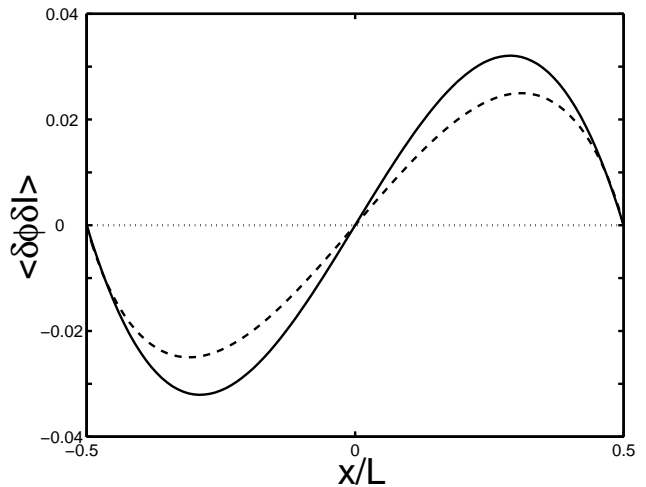}
 \caption{\label{fig:dphidI} The coordinate dependence of the zero-frequency
  correlator $\la\delta\phi(x)\delta I\ra_{\omega}$ normalized to $eV$ for the
  elastic case (solid line) and the hot-electron regime (dashed line).  }
\end{figure}
The correlator $\la\delta\phi(x)\delta I\ra$ vanishes at $V=0$ and is an odd
function of $x$ at nonzero $V$(see Fig.~\ref{fig:dphidI}).
\begin{figure}
 \epsfxsize=8.5cm
 \epsfbox{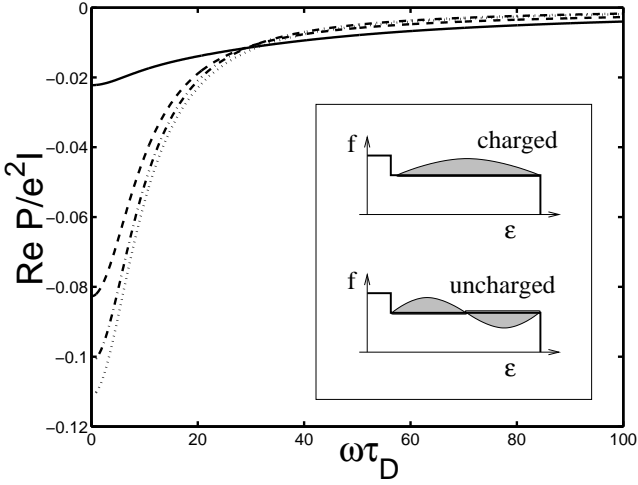}
 \caption{\label{fig:real} The real part of the ratio $P(\omega)/(e^2I)$
  versus normalized frequency $\omega\tau_D$ ($\tau_D = L^2/D$) for purely
  elastic scattering at $eV \gg T$ (solid line), hot-electron regime at $eV
  \gg T$ (dashed line), purely elastic scattering at $eV \ll T$ (dotted line),
  and hot-electron regime at $eV \ll T$ (dash-dotted line). Inset: charged and
  uncharged fluctuations of the distribution function $f$. Charged fluctuations
  have a short relaxation time and contribute to fluctuations of current $\delta I$.
  Uncharged fluctuations do not contribute to $\delta I$ directly but affect the
  intensity of noise sources. They decay only via slow diffusion and result in the
  low-frequency dispersion of the third cumulant.}
\end{figure}
\begin{figure}
 \epsfxsize=8.5cm
 \epsfbox{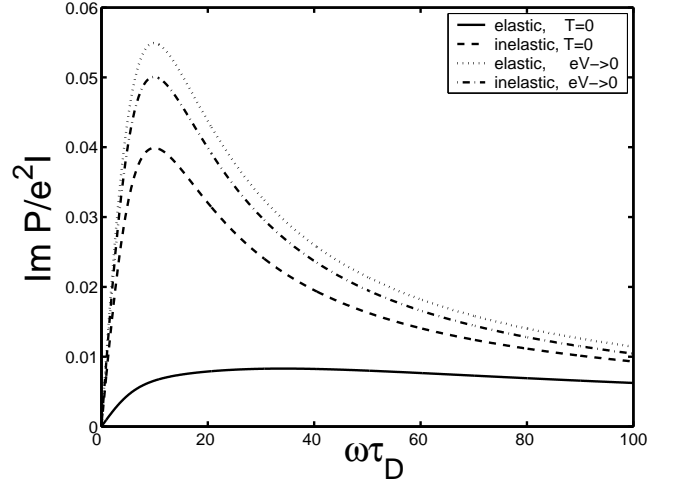}
 \caption{\label{fig:imag} The imaginary part of the ratio $P(\omega)/(e^2I)$
  versus normalized frequency $\omega\tau_D$ ($\tau_D = L^2/D$) for purely
  elastic scattering at $eV \gg T$ (solid line), hot-electron regime at $eV
  \gg T$ (dashed line), and purely elastic scattering at $eV \ll T$
  (dotted line), and hot-electron regime at $eV \ll T$ (dash-dotted line).  }
\end{figure}
Upon inverting the operator $(\nabla^2 + i\omega/D)$ in Eqs.
(\ref{<df1dI>}) and (\ref{<df2dI>}) and performing the spatial
integration in Eq. (\ref{P}), we arrive at an expression for
$P(\omega)$ in terms of $\qw = (i\omega/D)^{1/2}$, which is our
final goal. Because of its length, we give here only its
low-temperature and low-voltage limits
\begin{eqnarray}
 P_{el}(\omega)
 =
 -\frac{4}{3}
 \frac{e^2V}{R}
 \bigl[
  \qw L
  (\qw^2L^2 + 30)
  \sinh(\qw L)
\nonumber\\
  -
  8
  (\qw^2L^2 + 6)
  \cosh(\qw L)
  +
  2\qw^2 L^2
  +
  48
 \bigr]
\quad\nonumber\\
 \bigl/
 \bigl[
  \qw^5 L^5
  \sinh(\qw L)
 \bigr]
\qquad\qquad \label{P-zT}
\end{eqnarray}
and
\begin{equation}
 P_{el}(\omega)
 =
 \frac{4}{3}
 \frac{e^2V}{R}
 \frac{
  2
  \cosh(\qw L)
  -
  \qw L
  \sinh(\qw L)
  -
  2
 }{
  \qw^3 L^3
  \sinh(\qw L)
 }.
\label{P-zV}
\end{equation}
At $\omega = 0$ these expressions give $-(1/45)e^2V/R$ and
$-(1/9)e^2V/R$, which corresponds to $P_3(0,0) = -(1/15)e^2V/R$
and $P_3(0,0) = -(1/3)e^2V/R$. These zero-frequency results are in
agreement with Ref. \onlinecite{Lee-95,Gutman-03} . At finite
frequency, Eqs. (\ref{P-zT}) and (\ref{P-zV}) become
complex-valued and tend to zero as $i/\omega$ at
$\omega\to\infty$. The real and imaginary parts of $P(\omega)$ are
shown in Figs. \ref{fig:real} and \ref{fig:imag}.

\section{ Hot-electron limit }

Consider now the limit of strong electron-electron interaction.
In this case, the distribution function may be assumed to have a
Fermi shape with a coordinate-dependent temperature $T_e(x)$ and
electric potential $\phi(x)$
\begin{equation}
 f(\eps, x)
 =
 \left[
   1
   +
   \exp
   \left(
    \frac{ \eps - e\phi(x) }{ T_e(x) }
   \right)
 \right]^{-1}.
 \label{f-therm}
\end{equation}
If the frequency $\omega$ is smaller than the rate of
electron-electron collisions, a fluctuation $\delta f$ can be
expressed in terms of fluctuations of these quantities
\begin{equation}
 \delta f(\eps, \br, \omega)
 =
 \frac{
  \partial f(\eps, \br)
 }{
  \partial\phi
 }
 \delta\phi
 +
 \frac{
  \partial f(\eps, \br)
 }{
  \partial T_e
 }
 \delta T_e
 \label{df-hot}
\end{equation}
A substitution of Eq. (\ref{df-hot}) into Eq. (\ref{P}) and integration over
the energy readily gives
\begin{equation}
  P_{hot}(\omega)
  =
  \frac{2}{RL}
  \int\limits_{-L/2}^{L/2} dx
  \la
    \delta T_e(x)
    \delta I
  \ra_{\omega}.
  \label{P-hot-2}
\end{equation}
To calculate the correlator in Eq. (\ref{P-hot-2}), we have to
obtain a Langevin-type equation for $\delta T_e$. To this end, we
multiply Eq. (\ref{kinetic}) by $\eps$ and integrate it over
$\eps$, like it was done when deriving the equation of heat
balance in Refs.~\onlinecite{Nagaev-95,Kozub-95}. This gives
\begin{eqnarray}
 \left(
  \frac{\partial}{\partial t}
  -
  D\nabla^2
 \right)
 \left(
  \frac{\pi^3}{3}
  T_e\delta T_e
 \right)
 -
 D\nabla^2
 \left(
  e^2
  \phi\delta\phi
 \right)
\nonumber\\
 =
 -
 \int d\eps\,\eps
 \nabla\delta{\bf F}^{imp}.
 \label{Langevin-heat}
\end{eqnarray}
Multiplying Eqs. (\ref{Langevin-heat}) and (\ref{dI_L}) and
averaging the product with the help of Eq. (\ref{<dF^2>}) results
in an equation for the correlation function $\la \delta T_e \delta
I\ra_{\omega}$ of the form
\begin{eqnarray}
 \left(
  \nabla^2
  -
  i
  \frac{\omega}{D}
 \right)
 \left[
  \frac{\pi^2}{3}
  T_e
  \la
   \delta T_e(x)
   \delta I
  \ra_{\omega}
 \right]
 =
\qquad\qquad
\nonumber\\
 -
 \nabla^2
 \left[
  e^2\phi
  \la
   \delta\phi(x)
   \delta I
  \ra_{\omega}
 \right]
 +
 \frac{2e}{L}
 \frac{
  \partial
 }{
  \partial x
 }
 \int d\eps\,
 \eps f(1-f).
 \label{int-dI}
\end{eqnarray}
The integral over the energy on the right-hand side of Eq.
(\ref{int-dI}) equals $e\phi T_e$, and making use of Eq.
(\ref{<dphidI>-zf}), one easily obtains the solution of Eq.
(\ref{int-dI}) in a symbolic form
\begin{eqnarray}
 \la
  \delta T_e(x)
  \delta I
 \ra_{\omega}
 =
 \frac{6}{\pi^2}
 \frac{e^2}{L}
 \frac{1}{T_e}
 (
  \nabla^2
  -
  i\omega/D
 )^{-1}
\nonumber\\
 \times
 \Biggl\{
  \frac{
   \partial
   (
    \phi T_e
   )
  }{
   \partial x
  }
  -
  \nabla^2
  \left[
   \phi
   (
    \nabla^2
   )^{-1}
   \left(
    \frac{
     \partial T_e
    }{
     \partial x
    }
   \right)
  \right]
 \Biggr\},
 \label{dT_e-dI}
\end{eqnarray}
where the operators $\nabla^2$ and $\nabla^2 - i\omega/D$ are inverted with
zero boundary conditions. According to Ref. \onlinecite{Nagaev-95}, the mean
potential is given by $\bar{\phi}(x) = Vx/L$ and the mean effective
temperature by
\begin{equation}
 \bar{T}_e(x)
 =
 \left[
  T^2
  +
  \frac{3}{\pi^2}
  (eV)^2
  \left(
   \frac{1}{4}
   -
   \frac{x^2}{L^2}
  \right)
 \right]^{1/2}.
\end{equation}

As $\bar{\phi}$ and $\bar{T}_e$ are  odd and  even functions of
coordinate, the resulting correlator is an even function of $x$.
In the zero-frequency limit it is given by
\begin{eqnarray}
 \la
  \delta T_e(x)
  \delta I
 \ra_{\omega}
 =
 \frac{1}{2\pi^2}
 e^2 V
 \Biggl\{
  -
  a^2
  -
  2
  \frac{ x^2 }{ L^2 }
\nonumber\\
  +
  \frac{
   1
  }{
   \sqrt{ a^2 - 4x^2/L^2 }
  }
  \Biggl[
   (
    a^2 - 1
   )^{3/2}
   -
   3a
   \frac{x}{L}
   \arcsin
   \left(
    \frac{2x}{La}
   \right)
\nonumber\\
   +
   6
   \left[
    a^2
    \arcsin
    \left(
     \frac{1}{a}
    \right)
    +
    \sqrt{ a^2 - 1 }
   \right]
   \frac{ x^2 }{ L^2 }
  \Biggr]
 \Biggr\},
 \label{dT_e-dI-expl}
\end{eqnarray}
where $a^2 = 1 + (4\pi^2/3) T^2/(eV)^2$. This correlation function vanishes at
$V=0$ and is negative at positive voltages, i.e. if the increment of total
current through the contact is negative. This fact has a very simple physical
meaning: an increase in the total current results in an increase of the Joule
heating and hence an increase of the temperature.

An integration of Eq. (\ref{dT_e-dI}) with respect to $x$ gives
the low-frequency third cumulant for arbitrary temperatures and
voltages in a form
\begin{eqnarray}
 P_3(0,0)
 =
 \frac{3}{\pi^2}
 \frac{e^2V}{R}
 \Biggl[
  \frac{7}{12}
  -
  \frac{13}{4}
  a^2
  +
  \frac{1}{2}
  (
   5a^2 - 2
  )
  \sqrt{
   a^2 - 1
  }
\nonumber\\
  \times
  \arcsin
  \left(
   \frac{1}{a}
  \right)
  +
  \frac{1}{4}
  a^4
  \arcsin^2
  \left(
   \frac{1}{a}
  \right)
 \Biggr].
 \label{I^3-expl}
\end{eqnarray}
Its limiting values $P_3(0,0) = -(3/\pi^2)e^2V/R$ at $eV \ll T$
and $P_3(0,0) = -(8/\pi^2 - 9/16)e^2V/R$ at $eV \gg T$ coincide
with the results of Gutman and Gefen.\cite{Gutman-03}

In the case of nonzero frequencies, it is possible to obtain analytical
results only in the limiting cases of $eV \ll T$ and $eV \gg T$. In the
high-temperature limit, one may set $\bar{T}_e(x) = T$ in Eq. (\ref{dT_e-dI}),
hence $\phi(x)$ is the only coordinate-dependent quantity, and the term in
curly brackets equals $VT/L$. Then the diffusion operator is easily inverted
and the integration over $x$ gives
\begin{equation}
 P_{hot}(\omega)
 =
 -\frac{12}{\pi^2}
 \frac{e^2V}{R}
 \frac{1}{ q_{\omega}^2 L^2 }
 \left[
  1
  -
  \frac{2}{ q_{\omega} L}
  \tanh
  \left(
   \frac{ q_{\omega} L }{2}
  \right)
 \right].
 \label{P_hot-zV}
\end{equation}

In the zero-temperature limit, the term in curly brackets in
Eq. (\ref{dT_e-dI}) may be expanded in a Fourier series in $\cos[(2n+1)\pi
x/L]$ and the operator $\nabla^2 - i\omega/D$ is easily inverted. The final
result is obtained as a sum of an infinite series
\begin{eqnarray}
 P_{hot}(\omega)
 =
 \frac{12}{\pi}
 \frac{e^2V}{R}
 \sum\limits_{k=0}^{\infty}
\nonumber\\
 \times
 \frac{
   J_0(\pi k + \pi/2)
   [
    J_1(\pi k + \pi/2)
    -
    (-1)^k
   ]
 }{
   (2k + 1)
   [
    \pi^2
    (2k + 1)^2
    +
    i\omega
    L^2/D
   ]
 },
\label{series}
\end{eqnarray}
where $J_0$ and $J_1$ are Bessel functions of order 0 and 1.

\section{ Discussion}

Our results for the non-interacting regime (Eq. \ref{P-zT} and Eq. \ref{P-zV})
and for the hot-electron regime (Eq. \ref{P_hot-zV} and Eq. \ref{series}) are
displayed in the figures \ref{fig:real} and \ref{fig:imag}.  It is clearly
seen that both real and
 imaginary parts of the third cumulant have the most
pronounced dispersion in the
 case of a high temperature or for strong
electron-electron
 scattering, i.e. when the local distribution function has
a nearly
 Fermian shape. This unexpected result is in a sharp contrast with
the dispersion of quantum noise,\cite{Altshuler-94} which results
 from sharp
singularities in the energy dependence of the
 distribution
function. 

Mathematically, the different shape of the
 frequency dependence
for purely elastic scattering and interacting
 electrons at high voltages can
be explained as follows. In the
 case of hot electrons, both $\delta\la\delta
I^2\ra/\delta T_e(x)$
 and $\la\delta T_e(x)\delta I\ra_{\omega}$ are even
functions of
 the coordinate $x$ (measured from the middle of the
contact). On
 the contrary, for purely elastic scattering at $T=0$ both
$\delta\la\delta I^2\ra/\delta f(\eps,x)$ and $\la\delta
 f(\eps,x)\delta
I\ra_{\omega}$ are odd functions of $x$. The
 functions acted upon by the
inverse diffusion operator $(\nabla^2
 - i\omega/D)^{-1}$ in
Eqs. (\ref{dT_e-dI}) and (\ref{<df1dI>}) are
 also even and odd,
respectively. At low frequencies, the inverse
 diffusion operator is
essentially nonlocal in space and applying
 it results in an effective
averaging of the argument on the scale
 of the order of $L$. In the elastic
case, this averaging involves
 both negative and positive values, and this is
why the elastic
 third cumulant is suppressed at low frequencies as compared
to the
 hot-electron value. However at high frequencies, the inverse
diffusion operator becomes almost local in space, therefore there
 is no
averaging of negative and positive values and both cumulants
 become of
nearly the same magnitude. This absence of spatial
 averaging partially
compensates for the increasing frequency and
 makes the frequency dependence
of the "elastic" cumulant more
 flat. Therefore the different shape of the
frequency dependence in
 the elastic and hot-electron limits may be traced
back to the
 different symmetry of relevant second-order correlation
functions.
 
\section{ Summary }

 In summary, we have shown that diffusive contacts exhibit a
 nontrivial
internal dynamics even at frequencies much smaller than
 the inverse
charge-relaxation time. Though this dynamics is not
 affected by the electric
environment of the contact, it differs
 from the dynamics of charge-neutral
particles and manifests itself
 as a low-frequency dispersion of the third
cumulant of current.
 In view of the fact that both dynamic conductance and
shot noise of metallic conductors depend only on the $RC$ time, this frequency
dispersion of the third cumulant on the scale of the dwell time is a very
interesting result.

\begin{acknowledgments}
We thank B. Reulet and D. E. Prober for fruitful discussions.

This work was supported by the Swiss National Science Foundation, the program
for Materials and Novel Electronic Properties, the Russian Foundation for
Basic Research, Grant No. 01-02-17220, INTAS (project 0014, open call 2001),
and by the Russian Science Support Foundation.
\end{acknowledgments}

\appendix

\section{Stochastic Path Integral Representation}

In this appendix, we derive a stochastic path integral
representation~\cite{Pilgram-03,broader} for the full current
statistics of diffusive conductors. This representation then
serves to calculate the dispersion of the third order current
correlation function. We give first a detailed derivation of the
case without electron-electron scattering. For simplicity, we
restrict ourselves to quasi-one-dimensional wires and consider
only slow dynamics due to the longitudinal diffusion modes. In
this case, the kinetic transport equation for the fluctuating
contribution $\delta f(\epsilon,x,t)$ to the electron occupation
function $f=\bar{f}(\epsilon,x) + \delta f(\epsilon,x,t)$ is given
by the one-dimensional equivalent of Eq.~(\ref{kinetic}).  The
mean distribution function $\bar{f}$ is defined in
Eq.~(\ref{f-elast}). Since we restricted ourselves to the regime
of slow diffusive dynamics, we may assume charge neutrality
$\partial (\delta j) /
\partial x = 0$ where the fluctuations of the electrical current
density are given by Eqs.~(\ref{dj-vs-df}) and~(\ref{tot_curr}).
Finally, we make use of the fact\cite{Nagaev-02} that the
extraneous sources of noise are Gaussian random variables
delta-correlated in time, space and energy and described by
Eq.~(\ref{<dF^2>}).

In order to construct a path integral representation of this
Boltzmann-Langevin equation, we define the probability functional
$P$ which gives the probability to find a certain realization of
extraneous currents $\delta F^{imp}(\varepsilon,x,t)$. This
functional may be written as a path integral
\begin{eqnarray}
&P\left[ \delta F^{imp} \right] = &\nonumber\\
&\int {\cal D} \eta
\exp\left\{ \int dt  dx  d\varepsilon
(-i\eta\delta F^{imp} + H(\eta))
\right\}&
\label{Noise Probability}
\end{eqnarray}
taking Fourier transforms of the generating function
\begin{eqnarray}
H = -\frac{D}{N_F} f(1-f) \eta^2.
\end{eqnarray}
Since the Fourier transforms are independently taken for each point in space, time and
energy, Eq.~(\ref{Noise Probability}) indeed characterizes white noise. The electron
occupation function $f=\bar{f}+\delta f$ is considered to be a slowly changing variable
which modulates the instantaneous noise intensity. Its evolution is determined by the
kinetic equation~(\ref{kinetic}) which we represent as a delta functional expressed by a
path integral
\begin{eqnarray}
\label{Kinetic Constraint}
 \delta\left[ \delta \dot{f} - D\delta f''
 +
 e\frac{\partial f}{\partial \varepsilon}\delta\dot{\phi}
 +
 (\delta F^{imp})'\right]
 \\ \nonumber
 =
 \int {\cal D} \lambda
 \exp
 \Biggl\{
  i\int dt  dx  d\varepsilon\lambda
  \Bigl(
   \delta\dot{f}
   -
   D\delta f''
 \\ \nonumber
   +
   e\frac{\partial f}{\partial\varepsilon}
   \delta \dot{\phi}
   +
   (\delta F^{imp})'
  \Bigr)
 \Biggr\},
\end{eqnarray}
where the prime stands for $\partial/\partial x$. The dynamics of
the potential fluctuations $\delta\phi$ can be expressed by a
second delta functional which enforces charge neutrality
\begin{eqnarray}
\label{Electrostatic Constraint}
&\delta \left[ \sigma \delta \phi'' - eN_F \int d\varepsilon (\delta F^{imp})'
\right]
=
\int {\cal D} \xi&\\
\nonumber\\
&
\exp\left\{
i\int dtdx \xi\left(\sigma \delta \phi'' - eN_F \int d\varepsilon (\delta F^{imp})'\right)
\right\}.\nonumber
&
\end{eqnarray}
The fields $\lambda$ and $\xi$ can be understood as Lagrange multipliers.
Combining Eqs.~(\ref{Noise Probability}), (\ref{Kinetic Constraint}) and
(\ref{Electrostatic Constraint}), we construct the probability $P_t$ to find a
certain realization of extraneous currents $\delta F^{imp}$ under the
constraint of current conservation and charge neutrality
\begin{eqnarray}
P_t[\delta F^{imp}] =
\int {\cal D}\delta \phi {\cal D} \delta f
\delta[\dots]\delta[\dots]P[\delta F^{imp}].
\end{eqnarray}

We are now in a position to calculate the generating functional $S[i\chi]$
of current fluctuations $\delta I=\delta j(x=-L/2)$ at the left contact $x=-L/2$
\begin{eqnarray}
e^{S_{el}[i\chi]} =
\int {\cal D}\delta F^{imp} P_t[\delta F^{imp}]
\exp\left\{i\int dt \chi \delta I\right\}.
\label{Final Integrals}
\end{eqnarray}
This equation may be considerably simplified. In a first step, we
can integrate out the the extraneous currents $\delta F^{imp}$ as
well as the field $\eta$ introduced in Eq.~(\ref{Noise
Probability}). We are then left with four functional integrations
over $\delta f,\delta\phi,\lambda$ and $\xi$. In a second step, we
evaluate this integrations in the saddle point approximation. As
the diffusive conductor is essentially semiclassical, the
corrections to the saddle point action are
small~\cite{Pilgram-03}. After rescaling $\lambda \mapsto eN_F
\lambda$, we are left with the generating functional
\begin{eqnarray}
&S_{el}[\chi,\lambda,\xi,\delta f,\delta\phi] = \int dt dx \Bigl\{ \sigma
\xi\delta\phi'' + \int d\varepsilon \Bigl[ \frac{\sigma}{e}\lambda \delta f''+
&\nonumber\\
&
\sigma f(1-f)(\lambda'+\xi')^2
-eN_F\lambda\left(\delta \dot{f} + e \frac{\partial f}{\partial \varepsilon}
\delta \dot{\phi}\right) \Bigr] \Bigr\}&
\label{Cold Action}
\end{eqnarray}
which has to be evaluated at the saddle point given by
\begin{eqnarray}
\frac{\delta S_{el}}{\delta \lambda} = 0,\quad
\frac{\delta S_{el}}{\delta \xi} = 0,\nonumber\\
\frac{\delta S_{el}}{\delta (\delta f)} = 0,\quad
\frac{\delta S_{el}}{\delta (\delta \phi)} = 0.
\label{Saddle Equations}
\end{eqnarray}
Note that we performed a complex continuation
$i\lambda\mapsto\lambda,i\xi\mapsto\xi$ and $i\chi\mapsto\chi$.  We are
therefore left with purely real quantities.  The saddle point equations are
supplemented with boundary conditions: the three fields $\delta f,\delta
\phi,\lambda$ vanish at both boundaries. The external counting field $\chi$ is
incorporated into the boundary conditions for $\xi$
\begin{eqnarray}
\xi(-L/2) = \chi, \quad \xi(L/2) = 0.
\label{Boundary Condition}
\end{eqnarray}

The frequency dependent third cumulant under consideration in this paper is
obtained from the third functional derivative
\begin{eqnarray}
\langle \delta I(\omega_1)\delta I(\omega_2)\delta I(\omega_3)
\rangle =\left.\frac{\delta^3 S_{el}}{\delta \chi(\omega_1)\delta
\chi(\omega_2)\delta \chi(\omega_3)}\right|_{\chi=0}.
\end{eqnarray}
We calculate this cumulant by a systematic expansion of action~(\ref{Cold
  Action}), saddle point equations~(\ref{Saddle Equations}) and fields
\begin{eqnarray}
\delta f = \delta f_1 + \delta f_2 + \dots,\quad \delta\phi  = \delta\phi_1 +
\dots, \quad \dots
\end{eqnarray}
in orders of the external field $\chi$. It can be straightforwardly shown by
inserting saddle point equations~(\ref{Saddle Equations}) back into the
action~(\ref{Cold Action}) that the third order contribution to the action has
the form
\begin{eqnarray}
S_{el,3}[\chi] = \sigma \int dt \int dx \int d\varepsilon (1-2\bar{f})\delta f_1
(\xi_1')^2. \label{Third Action}
\end{eqnarray}
It remains to solve the saddle point equations to first order in $\chi$. For
the Lagrange multipliers we find
\begin{eqnarray}
\lambda_1''+\xi_1'' = 0, \text{ thus }
\lambda_1=0,\xi_1 = -\chi(x/L-1/2).
\end{eqnarray}
Their dynamics is trivial, since they follow instantaneously the
external field $\chi$. The interesting dispersion effect stems
from the saddle point equation for the occupation function which
takes the form of an inhomogeneous diffusion equation
\begin{eqnarray}
&\delta \dot{f}_1 - D \delta f_1'' = &\\
&-2eD\left(f_0(1-f_0) \xi_1'  \right)' -2eD\frac{\partial
f_0}{\partial \varepsilon} \int d\varepsilon \left(f_0(1-f_0)
\xi_1'  \right)'.&\nonumber
\end{eqnarray}
This diffusion equation has a very appealing interpretation: When we integrate
the equation over energy, we find that the two source terms on the right hand
side cancel.  The left side becomes a homogeneous diffusion equation for the
fluctuations of the charge density $\delta \rho_1= e\int d\varepsilon \delta
f_1$ which has the trivial solution $\delta \rho_1=0$. This is nothing than the
charge neutrality which we demanded in the beginning of this section.  The
second source term which is due to variations of the electrostatic potential
thus compensates the first term in  such  a way that all fluctuations of the
occupation function $\delta f_1$ are charge neutral. We decompose the total
variation $\delta f_1 = \delta f_1^F + \delta f_1^\phi$ into a contribution
$\delta f_1^F$ due to free fluctuations of the occupation function and a
contribution $\delta f_1^\phi$ due to potential fluctuations.  Using the
identity $\delta \dot{\phi}_1 = D \delta \phi_1''$, we solve the diffusion
equation by inverting the diffusion operator and find
\begin{eqnarray}
 \delta f_1^F(\omega)
 =
 \langle
  \delta f_F(\varepsilon,x)
  \delta I
 \rangle_\omega
 \frac{\chi(\omega)}{D},
 \\
 \delta f_1^\phi(\omega)
 =
 \langle
  \delta f_{\phi}(\varepsilon,x)
  \delta I
 \rangle_\omega
 \frac{\chi(\omega)}{D}.
\end{eqnarray}
The two correlators are defined in Eq.~(\ref{<df1dI>}) and
Eq.~(\ref{<df2dI>}) respectively. They can now be inserted into
Eq.~(\ref{Third Action}) to obtain the third-order correlation
function~(\ref{P}). We thus derived the cascade rules applied in
the main body of this paper from the stochastic path-integral
formalism.

The derivation of an action describing the hot electron regime requires only a
minor additional effort. Here we cite directly the result which has been
derived by one of the authors for the zero frequency limit in a different
context\cite{Pilgram-03b}
\begin{eqnarray}
& S_{hot} = \int dt\int dx \Biggl\{
-N_F \lambda T_e\dot{T}_e &\nonumber\\
& + \sigma \left(\xi'\quad\lambda'\right)
\hat{A}
\left(\begin{array}{c}\xi' \\ \lambda'\end{array}\right)
- \sigma \left(\xi'\quad\lambda'\right)
\hat{B}
\left(\begin{array}{c} \phi'\\ T_e'\end{array}\right)
\Biggr\}&
\label{Hot Action}
\end{eqnarray}
In this action, we introduced the local electron temperature
$T_e(x,t) = \bar{T}_e(x)+\delta T_e(x,t)$ and the local
electrostatic potential $\phi(x,t) = \bar{\phi(x)} + \delta
\phi(x,t)$. The boundary conditions for these fields are the
potentials and temperatures of the left and right reservoirs.  As
for the case of non-interacting electrons, the Lagrange multiplier
$\xi$ ensures charge neutrality and obeys the boundary
condition~(\ref{Boundary Condition}). The field $\lambda$ is
linked to the conservation of energy current and is zero at the
boundaries.  The matrix $\hat{A}$ describes the local noise
created by the extraneous sources of noise $\delta F^{imp}$
\begin{eqnarray}
\hat{A} = T_e\left(\begin{array}{cc} 1 & \phi \\ \phi
 & (\phi^2 + \pi^2T_e^2/3e^2)\end{array}\right).
\end{eqnarray}
The second matrix $\hat{B}$ is the linear response tensor
\begin{eqnarray}
\hat{B} =
\left(\begin{array}{cc} 1 & 0\\ \phi & \pi^2T_e/3e^2\end{array}\right).
\end{eqnarray}

In complete analogy to the derivation of Eq.~(\ref{Third Action}), we may
again collect all third order terms which contribute to the action~(\ref{Hot
  Action}) and find
\begin{eqnarray}
S_{hot,3}[\chi] = \sigma \int dt\int dx \delta T_{e,1} (\xi_1')^2.
\end{eqnarray}
where the variation $\delta T_{e,1}$ can be identified with
\begin{eqnarray}
\delta T_{e,1} = \la
  \delta T_e(x)
  \delta I
 \ra_{\omega}
\end{eqnarray}
(see Eq.~(\ref{dT_e-dI})). The total third order contribution
$S_{hot,3}[\chi]$ therefore corresponds exactly to Eq.~(\ref{P-hot-2}).

The main results of this appendix are the dynamic generating functionals
Eq. (\ref{Cold Action}) and Eq. (\ref{Hot Action}). These functionals permit
(in principle) the calculation of cumulants of arbitrary order.


\begin{thebibliography}{99}
\bibitem{footnote}
Our results can be generalized to the case of nonideal leads by
inserting them into the expressions of Refs. \onlinecite{Beenakker-03}
and \onlinecite{Kindermann-03}.
\end{thebibliography}
\end{document}